\documentclass[10pt,article]{aastex}
\usepackage{emulateapj5,natbib}

\newcommand{\gtsim}{\mbox
{{\raisebox{-0.4ex}{$\stackrel{>}{{\scriptstyle\sim}}$}}}}

\newcommand{\tm}{2.2 $\mu $m }

\shorttitle{IR Synchrotron Emission in Cas A}
\shortauthors{Jones et al.}
\begin{document}
\title{The Identification of Infrared Synchrotron Radiation from 
Cassiopeia A}
\vspace{0.4cm}
\author{T. J. Jones\altaffilmark{1}, L. Rudnick, T. DeLaney \& J. Bowden }
\affil{Astronomy Department, University of Minnesota,
    Minneapolis, MN 55455}
\vspace{0.5cm}
\altaffiltext{1}{ Visiting Astronomer at the Infrared Telescope Facility,
which is operated by the University of Hawaii under contract from the 
National Aeronautics and Space Administration.}
\begin{abstract}
    We report the discovery of polarized flux at \tm from the bright
    shell of the $\approx$ 320 year old supernova remnant Cas~ A. The
    fractional polarizations are comparable at 6 cm and \tm, and the
    polarization angles are similar, demonstrating that synchrotron
    radiation from the same relativistic plasma is being observed at
    these widely separated wavebands.  The relativistic electrons
    radiating at \tm have an energy of $\approx 150~$ GeV, ($\gamma
    \approx 3\times 10^{5}$), assuming an $\approx 500~ \mu$G magnetic
    field.  The total intensity at \tm lies close to the power law
    extrapolation from radio frequencies, showing that relativistic particle
    acceleration is likely an ongoing process; the infrared emitting
    electrons were accelerated  no longer than $\approx$  80
    years ago.  There is a small but significant concave curvature to
    the spectrum, as expected if the accelerating shocks have been
    modified by the back pressure of the cosmic rays; given 
    calibration uncertainties, this conclusion must be considered 
    tentative at present.  The \tm
    polarization angles and the emission-line filaments observed by
    HST are both offset from the local radial direction by 10 $-$ 20
    degrees, providing evidence that the magnetic fields in Cas~A are
    generated by Rayleigh-Taylor instabilities in the decelerating
    ejecta.
    
\end{abstract}

\section{Introduction}
\label{sec:intro}
The energy distribution of $\approx 0.1 - 10$ GeV electrons in
supernova remnants has been studied for decades through their radio
synchrotron radiation.  The distributions are typically power laws,
with low energy spectral flattenings often seen due to absorption in
the interstellar medium \citep{Kas89}, or perhaps even due to
internal ionized gas in the case of Cas~A \citep{Kas95}.  The power
law slopes (spectral indices) vary significantly from remnant to
remnant  \citep{Gre01}.  Although it has long been accepted
that diffusive shock acceleration is responsible for the GeV electrons
\citep{Bel78}, there are still no clear observational signatures for
how this is regulated in different remnants.  Spectral index
variations within an individual remnant are often very small; this
presents a dilemma because the asymmetries in remnant structures and
dynamics should lead to large variations in spectral index if particle
acceleration is due to first-order Fermi acceleration at shocks.  In
older remnants, such as the Cygnus Loop, spectral variations probably
reflect local differences in absorption and confusion along the line of
sight \citep{Lea98}.  But in Cas~A, the spectral variations across
the remnant \citep{And96} are most consistent with different power
laws, indicating local variations in relativistic particle
acceleration \citep{Wri99}.  Information about deviations from the
power law shape could provide important information about the
acceleration mechanisms and the possible modifications of the shocks
by cosmic-ray backpressures  \citep{Ber99}; only weak
evidence for such deviations exists to date \citep{Rey92}.

The maximum energy to which cosmic rays can be accelerated in
supernova remnants is another issue of current importance. 
Non-thermal X-ray synchrotron radiation from the supernova remnants
SN1006 \citep{Koy95, Rey96}, RX J1713.7-3946 \citep{Koy97,Sla99}, Cas
A \citep{All97}, and RCW86 \citep{Bor01} suggest that relativistic
electrons have energies up to 100 TeV. Gamma-ray detections of SN1006
\citep{Tan98} and RX J1713.7-3946 \citep{Mur00}  have 
been interpreted as the Inverse Compton emission from these high 
energy electrons against the cosmic microwave background.

Although the energies of electrons that would produce synchrotron
radiation in the infrared are considerably lower than those
emitting  X-rays and
$\gamma-$rays, they offer the possibility of a
model-independent probe of electrons at much higher energies than seen
in the radio.  The possibility of near or mid-IR synchrotron emission
was first suggested to us by R. Tuffs (private communication and \cite{Tuf97}), based
on their impressions of the ISOCAM images \citep{Lag96}.  The
possibility was again raised by \citet{Ger01}, who noted the close
similarity between their \tm (but not 1.2 $\mu $m) image and the radio
images.  This led us to seek verification of the infrared synchrotron
origin through polarization measurements.

\section{Observations}

\subsection{Choice of Field}

The observations were made using NSFCAM on the IRTF in polarimetry
mode on 1 August, 2001.  Since NSFCAM has a field of view of about
75$\arcsec$ (0.3$\arcsec$
 per pixel) and we would have time to observe only one
field, we had to make a careful choice of location within the Cas A
remnant.  The \tm  band image kindly provided by C. Gerardy and R. Fesen
 \citep{Ger01} gave us information on the flux from Cas A at \tm.
  Our own radio images of Cas A  \citep{And96,And95,Goe01}
 provided us with total intensity and polarized
intensity maps that could be compared to the \tm band image.

Our choice of which region in Cas A to observe was influenced by a
number of factors.  First, there had to be a good correspondence
between the radio synchrotron emission and the \tm emission.  This was
done to avoid small bright regions of emission at \tm that might be
contaminated by emission lines unrelated to the radio synchrotron
emission.  Second, there had to be significant polarized intensity in
the radio, which we hypothesized would correspond to similar
polarization in the near infrared.  Third, there had to be sufficient
flux at \tm to make fractional polarization measurements of a few
percent, a level similar to the observed radio polarization.  Using
these criteria, we chose a segment of Cas A's  bright ring in the
northwest (azimuths of $\approx 0 to  -40$ degrees) centered at  RA 23h 23m 26s , 
Dec 58$^{\circ}$,  48$\arcmin$ 58.4$\arcsec$, (J2000),
 as shown in Figure 1.

\subsection{Polarization Observations}

The techniques for using NSFCAM as an imaging polarimeter are
described in \citet{Jon97,Jon00}.  Images are taken of the source and
an off-source sky location at four positions of a half waveplate that
can be rotated in the beam.  This allows us to form sky subtracted
intensity images at polarization position angles of 0, 45, 90 and
135$^\circ$ (waveplate angles of 0, 22.5, 45 and 67.5$^\circ$). We
formed two independent estimates of the total intensity ($I_{0}+I_{90}$)
and ($I_{45}+I_{135}$), and found that these were consistent with each
other within
the off-source noise errors.  Photometric calibration was done 
using observations of
HD 225023 from the Elias standards list \citep{Eli82}.  The \tm (K)
intensity scale was set using the conversion 
$0.0~ m_K = 635$ Jy.  The
final accuracy of our spectral indices is limited by the \tm
extinction corrections, as discussed below.

 From
the images at four angles, we then 
computed Stokes images Q ($I_{0}-I_{90}$) and U
($I_{45}-I_{135}$) and the total intensity image $\frac{1}{2}
(I_{0}+I_{45}+I_{90}+I_{135})$.  Note that these Stokes parameter
definitions differ by a factor of two from those commonly used in the
radio.  The details of our sequencing of source, sky and waveplate
position differed somewhat from \citet{Jon97}.  For Cas A we rotated
back and forth between two waveplate positions (either for Q or U) for
several minutes, then repeated the same procedure on the sky position. 
Integrations at a single waveplate position were 10 seconds.  It took
about 2 seconds for the waveplate to move to a new position.

NSFCAM is not a true imaging polarimeter, but rather a camera
retrofitted with some polarization optics, so we were limited by
systematics in the precision of our polarimetry to about $\pm$0.3\%. 
This limit is due primarily to fluctuations in atmospheric
transmission between the time it takes to integrate at different
positions of the waveplate.  Since we are interested in polarizations
of a few percent or more in Cas~A, this instrumental limitation did
not hinder our observations.  In fact, the nebulosity in Cas A at
\tm is sufficiently faint that our measurements are effectively
photon noise limited by the background.

Calibration of the position angle was done using observations of AFGL
490 two nights earlier with the same observing procedure as for Cas A.
AFGL 490 was assumed to have a polarization angle of $\theta$ =
115$^\circ$ at \tm \citep{Kob78}. The Q' and U' values quoted in
Table 1 are given as observed;  the PA contains a $-33$ degree correction
to place it in the proper sky frame.   An extra check on the position
angle calibration was done using observations on 1 August of the
reflection nebulosity surrounding the young star RNO1 \citep{Wei93}. 
The centro-symmetric scattering pattern around this star produces all
possible position angles and provides a very sensitive calibration. 
Calibration of the efficiency (which exceeds 98\% at \tm) was
performed in previous observing runs using S1 in $\rho$ Oph which has
a polarization of 1.9\% at \tm. Instrumental polarization was checked
during the night by observing an unpolarized standard star from the
UKIRT faint standards list (FS 149) and was found to be too small to
be measured.

The telescope was offset 100$\arcsec$ to the North for the sky frames.  This
was intended to move the telescope well off Cas A, but not so far that
telescope movement and settling time seriously reduced our observing
efficiency.  In an effort to increase our time on Cas A itself, we
reduced the number of background frames to about half the number of
frames taken on the source. The total on source integration time
was $\approx 1$ hr. Serious telescope drift during the long
integrations on Cas A resulted in the movement of stars in the
background frames relative to the source frame.  This produced
elongated stars in the background subtracted images (see Figure 1, \tm
intensity).  No correction has been made for these problems, and the
effect on our spectral measurements is described below.  Our
polarization measurements are less affected by the star trails than
the total intensity because the star polarizations are expected to be
$\approx 1 \% $ as discussed below, while Cas A's polarization is
$\approx 5 - 10 \% $.

Since the time between source and background frames ranged over
several minutes, fluctuations in sky background could cause a residual
DC level to be present in the sky subtracted images.  Although small
compared to the net intensity of the brighter nebulosity in Cas A,
this DC offset could be comparable to the polarized intensity, and
produce a spurious net intensity in the Q and U images.  These offsets
were removed by first identifying several small regions of the
sky-subtracted total intensity image with values fluctuating near zero
for the intensity.  We simply assume the true intensity is zero when
averaged over these locations and subtract the respective means from
the I, Q and U images to force them to be zero there.  This procedure
will remove the DC offset produced by sky background fluctuations in
the Q and U images, but also would remove any real, but faint emission. 
Sensitive observations in the radio do show faint total intensity and
polarized emission from these regions, at intensities of approximately
15\% that of the bright ring; no correction has been made for this
small bias.

\subsection{Radio and Optical Images}

The radio images were produced from observations at the Very Large
Array \footnote{The Very Large Array is an instrument of the National 
Radio Astronomy Observatory, a National Science Foundation facility 
operated under cooperative agreement by Associated Universities, Inc.}
 using all four
configurations over the time period from May, 1997 through March,
1998.  Maps were made by using the data from 1.285 GHz for the 20 cm
band, and from 4.64 GHz for the 6 cm band.  The flux calibration scale
was set using standard VLA procedures \citep{NRA02} and the source
3C48.  The polarization calibration also used standard VLA procedures
with the angle fixed by assuming a position angle of $-$10$^{\circ}$ for
3C138.  Other details of analysis and image creation are similar to
those described by \citet{And95}.  The full resolution of these images
was $\approx 1.5 \arcsec$.

The optical emission in Cas~A's bright ring is dominated by clumpy
emission lines on a wide range of scales, most recently studied in
great detail by \citet{FesHST01}.  The emission line material
represents a distinct temperature component, separate from the
synchrotron plasma, but which may dominate the dynamics driving the
relativistic particles and magnetic fields.  In particular,
\cite{FesHST01} noted from their high resolution WFPC2 observations, a
very large number of filamentary structures that they suggested were
Rayleigh-Taylor instabilities.  Although there is significant
small-scale scatter, we measured the approximate position angle of the
filaments at the azimuths of the  boxes used for the  synchrotron
 measurements and list these in Table 1.

\section{Data Analysis \& Results}

   \subsection{Image Registration} The registration of the \tm and
   radio images was done in two steps.  The radio coordinates from the
   VLA are already in the FK4 reference frame to an accuracy of $\approx 0.1
   \arcsec$.  To put the \tm image in the same frame, we first used
   the Sloan Digital Sky Survey to register the image of the entire
   Cas~A field kindly provided by C. Gerardy \& R. Fesen \citep{Ger01}.  We
   then minimized the rms differences between our new small
   polarimetry field and the Gerardy/Fesen image, allowing for shifts
   in position, rotation, and scale.  The final accuracy of this
   registration is $< 0.1 \arcsec$, and the uncertainties have no
   effect on the results discussed here.
 
\subsection { Spatial averaging} The signal:noise on the $0.3
\arcsec$ pixel level was not sufficient either for accurate spectral
measurements or determination of the polarization.  We therefore found
the average brightness for I, Q, and U in each waveband in six boxes
as shown in Figure 1, and listed in Table 1.  These boxes were chosen
to have sufficiently high polarized flux at 6 cm, to avoid regions
with bright stars in the sky (``off'') frame, and to avoid the edges of
the frame where the flat fielding is less accurate.
   
   For display purposes, we also convolved the polarizations (Q and U)
   to a resolution of 10$\arcsec$.  The resulting low resolution
   polarized intensities are shown on the right side of Figure 1.  The
   middle image of the right side represents a prediction of the \tm
   polarizations using the radio data and \tm total intensities.  It 
   was constructed by taking the fractional polarization image at 6 cm 
   and multiplying it by the total intensity image at \tm.  The overall
   correspondence is good, but not perfect.  The \tm polarized 
   intensity map contains, e.g., a bright spot due to residuals from 
   the bright star immediately to the SE of box 6.  In addition, it 
   has not been corrected for the noise bias, which is responsible 
   for some of the structure in the image.  The 6 cm polarization 
   angles were used, after correction for Faraday rotation as 
   described below, as a prediction for the \tm angles.  The 
   correspondence here is somewhat better than expected from the 
   errors. 

   A number of corrections to the total and polarized intensities are
   needed before the \tm and radio intensities and polarizations can
   be quantitatively compared.  First is the contamination of the \tm
   total intensity by foreground stars in the boxes.  The fluxes of
   the brightest stars in each box were therefore subtracted yielding
   the ``corrected'' \tm intensities given in Table 1.  We estimate
   that the remaining positive star contamination is less than a few
   percent.  
   
   More troublesome are the negative trails from stars in the sky
   frame, since the contribution from each individual star is spread
   across the box, and is therefore fainter and harder to recognize. 
   Although the boxes avoid the brightest negative trails, there are
   certainly unrecognized trails that reduce the \tm intensities.  In
   individual regions, such as the NW corner of the frame, we can look
   for negative trails in the absence of bright diffuse emission from
   Cas~A. The distribution of pixel values in that region is
   consistent with a gaussian distribution around mean zero with a
   long negative tail.  For this region, the negative tail can be
   recognized and ignored in determining the zero offset level.  When
   negative star trails overlap bright \tm emission, however, they can
   reduce the intensity without being recognized.  There is thus a
   negative bias in the \tm intensities, which we expect to be small
   on average ($\approx 0.5 \mu$Jy/$\sq \arcsec$, or $\approx 5\% $ of
   the total intensity), but could possibly be as much as $\approx 30\%
   $ in an individual box, without being recognized.  The \tm total
   intensity values quoted in Table 1 could therefore actually be
   somewhat larger, but because of the scientific significance of such
   larger values, we have taken a more conservative approach and not
   tried to make any corrections for this possible bias.
   
   Another bias correction that we did {\it not} make to the data is
   due to the fact that the sensitive radio observations show that
   synchrotron emission is present {\it everywhere} in the \tm image,
   and therefore it is not really possible to define an appropriate
   zero, as discussed earlier.  The characteristic intensity of the
   radio plateau is approximately 15\% of the bright ring, so defining
   it to be zero at \tm results in spectral indices between \tm and
   6 cm that are too steep by $\approx 0.01$.
   
   Cas~A's high optical extinction ($> 4^{m}$ \citet{Sea71} ) requires
   that corrections are needed even at \tm.  \citet{Hur96} find
   reddenings of $4.6 < A_{V} < 5.4$ magnitudes for five FMKs and
   $A_{V}$ = 5.3 and 6.2 magnitudes for two QSFs, and provide an
   extensive discussion of the literature on Cas~A's extinction.  In
   the west, $A_{V}$ may extend up to 8 magnitudes, according to the
   CO and HI analysis of \citet{Tro85}.  In this paper, we use  a
   value of $A_{V}$ = 5 magnitudes, which
   leads to $A_{2.2\mu m}$ = 0.55 magnitudes, adopting the conversion
   factors of \citet{Bes98}.  Application of this extinction value
   results in the lines labeled ``de-reddened'' for I, Q, and U in
   Table 1, and the values plotted in Figure 2.  We also show, in
   Figure 2, the range of corrections that would result for the range
   $4.6 < A_{V}€ < 6.2$ magnitudes.

   For any reasonable value of the extinction, it thus appears that
the \tm data consistently fall above the extrapolation from radio
wavelengths, giving the spectra a concave shape, i.e., flatter at
short wavelengths. There are several caveats to this conclusion.
First, a mixture of different power-laws, as observed at cm wavelengths
\citep{And96, Wri99}, will result in a concave spectrum.  The rms
variations within our boxes are $<= 0.01$, insufficient to cause a
significant effect at \tm.  However, it is likely that there is
some contamination of the bright ring with emission from the
steeper spectrum plateau, at least at radio wavelengths. The maximum
contamination, for plateau emission with $\alpha^{20cm}_{6cm}=-0.9$
would result in an increase of the \tm flux density of $\approx 35\%$.
The actual \tm contamination is likely to be considerably lower, given
our procedure for setting a zero background, as discussed earlier.  However,
in any case, this possible contamination is much less than the factor
of two enhancement seen at \tm after correction for extinction. 

 A second concern is our calibration to
 the radio flux scale \citep{NRA02}, which itself is
 based on ``absolute'' calibrations of Cas A \citep{Baa77}.
These two steps are expected to be accurate to $\approx 2\%$, and 
so could each
contribute another $25 \%$ error to the \tm extrapolation.  We
therefore consider our detection of spectral concavity to be reasonable,
but somewhat tentative until the extinction and flux scale issues can
be fully resolved. At present, we are analyzing the 6 cm $-$ \tm spectra 
across the whole remnant, which should allow us to use differential
measurements to  separate real from calibration  effects. 


   \subsection{Polarization}  The \tm  Stokes parameter
   values in Table 1 are listed as 
   Q$^{\prime}$, U$^{\prime}$;
 these represent the actual observations, which are only 
   nominally Q and U prior to the calibration of the absolute 
   polarization angle.  The properly calibrated polarization position 
   angles are listed in Table 1.
   
       In either the (Q$^{\prime}$,U$^{\prime}$)
 or (Q,U) frame, the observed polarized
       intensity ($\sqrt(Q^{2} + U^{2})$) represents a biased measure
       since it would give a positive result from noise, even in the
       absence of a true signal.  As a simple correction for this
       bias, the fractional polarizations for \tm listed in Table 1
       were calculated as $\% P = [\sqrt(Q^{2}- \sigma_{Q}^{2}+
       U^{2}-\sigma_{U}^{2})]/I $ where $\sigma_{Q,U}$ were calculated
       as the rms scatter among boxes of approximately 35 pixels on a
       side, away from bright stars and away from the bright ring.
       
For the \tm polarization measurements, we also need to consider the
possible contamination by the foreground interstellar polarization
expected from the large extinction towards Cas~A. Catalogs of optical
polarimetry (e.g. \citet{Hei00}) have no stars in the same field as
Cas~A, so a direct measurement of the interstellar polarization in the
visual has not been made.   We have  measured the polarization of the
bright star in the northwest of our image and find a 3$\sigma$ upper
limit of 0.9\% at \tm, although we do not know if it is as far
away as Cas~A. The expected \tm polarization for Cas~A, given the
visual extinction of 5-6 magnitudes, can be estimated using the mean
trend in polarization with extinction analyzed in \citet{Jon89} and
\citet{Jon91} of $$<P_{2.2\mu m} (\%)> = 2.2*\tau_{2.2\mu m} = 0.2*A_V
\approx 1 \% .$$   Thus, we conclude that interstellar polarization is
not a significant contaminant of our polarimetry of the nebulosity at
\tm. After consideration of all these effects, and the faintness of the
nebulosity being measured, the errors in the fractional polarizations
and angles at \tm are dominated by the random contribution from sky
background photons, as opposed to systematic effects.   The
measured percentage  polarizations at \tm range from $\approx 4 - 10$ \%,
comparable to those at 6 cm.


To compare the \tm polarization angles to those at radio wavelengths,
we must first correct for  Faraday Rotation, which can arise
 both local to Cas~A and along
the line of sight. 
At 20 cm, the rotation is strong enough to actually depolarize the
emission \citep{And95}, and render rotation measure (RM) measurements
at that wavelength unreliable for the higher frequencies. 
\citet{Ken85} present RM measurements at low resolution but high
frequencies, with a value of $ -106~ \rm{rad~ m^{-2}}$ for the northwestern
part of the remnant.  In order to look for variations in RM among the
boxes used here, we created three different polarization maps at 4.41, 4.64 and
4.99 GHz, using the VLA D configuration data alone to produce a beam
size of $\approx 10 \arcsec$.  The signal:noise was insufficient to
detect significant variations among the boxes, but averaging over all
boxes yields $RM = -105 \pm 3 ~ \rm{rad~ m^{-2}}$, consistent with the
\citet{Ken85} value.  We therefore adopt the value of
 $-106 ~ \rm{rad~ m^{-2}}$
for all boxes, and show the  Faraday corrected position angles in Figure
1 and in the angle comparisons at the bottom of Table 1.

In Figure 3, we compare the position angles of the magnetic fields
derived from the polarization vectors in each box with the approximate
angles of the filamentary structures in the WFPC2 images of
\citet{FesHST01}.  All of the position angle measures change
systematically with azimuth, with a small but significant clockwise
offset from the local radial direction at least 
for the \tm and filament angles.  It is not clear if the 6 cm
angles show any offset, but better local rotation measures would
be needed to make the comparison with \tm angles more accurately.

%

\section {Scientific Implications}
\subsection{Maximum energy of accelerated electrons}
Any mechanism to accelerate relativistic particles yields some maximum
particle energy.  The energy can be limited by a variety of factors
including the time available for acceleration, the escape of the
highest energy particles from the finite sized acceleration region, or
the balance between energy gains and losses \citep{Rey98,Rey01}.
  In Cas~A, a hard X-ray tail, which falls two orders of magnitude below
the flux extrapolated from radio wavelengths, suggests that the
synchrotron emission may extend to that regime \citep{All97}.  This
conclusion has been called into question by \citet{Ble01}, who argue
that the 10 $-$ 15 $keV$ emission is diffuse, and not concentrated at
the outer shock as it should be if it were synchrotron.  This argument
is questionable, however, because the radio synchrotron radiation also
does not show an enhancement at the location of the outer shock; Xray
synchrotron emission is therefore still a possibility.  Bremsstrahlung
provides an alternative explanation for the hard X-ray emission
\citep{Lam01}, where lower hybrid waves in Cas~A's strong magnetic
field scatter electrons up to energies of tens of keV. There is not
yet any good test to distinguish between bremsstrahlung and
synchrotron explanations for Cas~A, so the nature of the 10~-~100
$keV$ emission is therefore still open to question.  With the
demonstration of a synchrotron origin for the \tm emission, it is
therefore interesting to estimate the maximum energy of the
accelerated particles.

We start by calculating the value of the field strength that minimizes
the energy in both relativistic particles and magnetic field
($B_{min}$).  We assume a pathlength through the emitting regions of
0.5 pc, with a filling factor of unity.  We assume no energy in
relativistic protons.  These assumptions lead to a 
 field strength of $\approx 500~ \mu$G for the regions in
Table 1.   This field strength can then be used to
estimate the energy of the relativistic electrons that are radiating
at \tm .  In a $\approx 500~ \mu$G field, this leads to energies of
$\approx 150$ GeV, or a relativistic gamma of $3 \times 10^{5}$. As
argued below, the synchrotron spectrum almost certainly extends a
factor of 10 higher in frequency, which would result in particle
energies(gammas) of $\approx 450$ GeV ($9 \times 10^{5}$). 
This is still a factor of
at least $10^{3}$  below energies to which SNRs are commonly expected to
accelerate cosmic rays, but it is the first unambiguous demonstration of
the extension of the same electron population to such high energies in
a shell SNR.

\subsection{Synchrotron loss limits}
We use the same field strength calculation, and the lack of spectral
steepening between 20 cm and \tm to place a limit on the time since the
relativistic particles were last accelerated.  The assumptions in the
field strength calculation bias the lifetimes to longer values, so our
lifetime estimates should be considered upper limits.  At a factor of
10 in frequency below the ``break'' frequency, the spectrum should
already have steepened by 0.1 (e.g., \citet{Lea91}, Figure 3.3a). 
Since, if anything, the spectra have actually flattened at high
frequencies, we adopt 0.1 as a conservative upper limit to the
steepening at \tm, yielding a ``break'' wavelength of $< 0.22\mu$m. 
This leads to  lifetimes for the electrons radiating in the infrared
that are  no longer than 80 years. It is thus likely that particle
acceleration is  an ongoing process, at the current stage of
Cas A's evolution.  Based on
our latest dynamical picture, this is the period when Cas~A has
probably swept up $\approx 1 - 10$ times
 its own mass \citep{Goe01,Wil02,Del02} and
during which the reverse shock is slowly coming to the end of its
outward motion (in the sky frame).  The reverse shock is still strong
at this stage, and is encountering outward moving ejecta at speeds of
$\approx 5000-6000$ km/s \citep{Ree95}. The x-ray emitting
filaments  are  rapidly decelerated to 
$\approx 3500$ km/s while those in the radio are slowed to
$\approx  1800$ km/s \citep{Kor98, Vin98, Del02};  this  provides at least one possible source of
energy for the relativistic particle acceleration.

\subsection{Non-linear shocks}
If the acceleration of cosmic rays at SNR shocks is efficient, then
there is a back reaction of the cosmic rays on the shock strength and
structure \citep{Dru81}. When the maximum cosmic ray momentum is limited
by shock properties such as geometry, the compression ratio can become
arbitrarily large \citep{Eic84}. The increasing gyro-radii at
higher energies allows those particles to see a higher compression due
to the sub-shock structure, so the resulting spectrum
becomes concave \citep{Bel87,Ber99,Rey92}, as observed here.  The radio
spectra can then be much steeper than the $-$0.5 expected in the strong
shock test particle limit, and can flatten out considerably before
they cut off, e.g., in the X-ray, due to radiative losses and escape.
\citet{Rey92} suggested that slight curvature might have been
observed in the integrated radio spectra of Tycho and Kepler SNRs,
although the signal was marginal.

In this work, we have shown that the \tm data, after correction for
extinction, fall significantly above the power law extrapolation from the
radio.  The mean flattening between $\alpha^{6cm}_{20cm}$ and
$\alpha^{2.2\mu m}_{ 6cm}$ is 0.06 (Table 1), equivalent to higher fluxes
at \tm by a factor of $\approx 2$ .  If the curvature were constant
across this full range of wavelengths, then the {\bf local} spectral
index would change from $\approx -0.75$ at cm wavelengths
 to $\approx -0.5$ at \tm.  This is essentially the same as the model shown by
\citet{Ber99}.  Using their Figure 5, we made a rough calculation of
the flux expected at \tm for Box 1; this is shown as the nested square
symbol in Figure 2.  The agreement is quite good, suggesting that the
signature of acceleration at modified shocks has been observed.  In
the context of the \citet{Ber99} model, this implies a critical
injection rate of relativistic particles of $\gtsim 10^{-4}$ and, for
most of the parameters they explored, efficiencies of 10 $-$ 100 \%. 
Thus, with the caveats described earlier regarding the extrapolation
of the radio spectrum,
 relativistic particles appear to be an important part of the
dynamics of the shock and the overall evolution of the SNR.

\subsection{Polarization}
The fractional polarizations and the polarization angles are very
similar at 6 cm and \tm .  This  provides confirming evidence not only
for synchrotron radiation, but that we are looking at the same
population of electrons at both wavelengths, validating the detailed
spectral comparisons above.  The \tm results confirm the expectation
from the \citet{Ken85} mm wave studies that fractional polarizations of
$\approx 5-10 \%$ represent real field disorder, as opposed to Faraday
depolarization.  Thus, although we speak about the ``radial'' magnetic
fields in Cas~A, this only refers to the small net ordered component 
field.  Approximately half of the field energy must be in a component
which is disordered on scales $< 1 \arcsec$ ($<0.02$ pc at a distance
for Cas~A of 3.4 kpc).

\citet{Gul75} first suggested that Rayleigh-Taylor instabilities in
the decelerating bright ring of Cas~A and other young SNRs would lead
to radial magnetic fields.  Detailed numerical simulations since that
time confirm the prevalence of these filamentary structures and their
role in the amplification and alignment of the magnetic fields
\citep{Jun96,Jun99}.  Approximately radial filamentary
structures have been seen in the Crab SNR, and, more recently in the
WFPC2 images of Cas~A \citep{FesHST01}.  In the northwest region
investigated here and shown in \citet{FesHST01} Figure 8, the Cas~A
filaments emerge from the suggested location of the reverse shock in
directions that are systematically non-radial.  These non-radial
motions could arise from large scale asymmetries in the ejecta, as is
well established from the distribution of different elements seen in
the X-ray (e.g. \citet{Hwa00}).  Alternatively, gradients in the
circumstellar medium could cause asymmetric deceleration (e.g.,
\citet{Kor98}), again leading to large scale velocity shears.  In any
case, the close correspondence between the R-T finger orientations and
the local magnetic field directions strongly supports a connection
between the two.  \citet{Blond01} show that the R-T
instabilities generated at the reverse shock can extend all the way
out to the outer shock front, providing radial magnetic fields instead
of the tangential ones otherwise expected at the shock.  The
transition region between the radial and tangential regions may have
been observed in Cas~A \citep{Goe01}, although the signal-to-noise is
quite low.

\section {Conclusions}

The extension of Cassiopeia A's known synchrotron spectrum to \tm
opens up a variety of important issues regarding the acceleration of
relativistic particles. First, we have established that the 
acceleration mechanism, usually assumed to be diffusive acceleration
at shocks, operates to energies of at least 150 GeV in Cas A.
In addition, there is reasonable evidence that these shocks are
efficient particle accelerators, arising from non-linear interactions
between the particles and the shocks.  This creates the opportunity
to investigate these interactions in conjunction with the dynamical
information about the several coupled plasmas (thermal and
non-thermal) in the remnant.  In turn, this will strongly constrain
models for the high X-ray energy tail in Cas~A's spectrum and assist
in the challenge of isolating the energy in relativistic protons.  To
make progress on these issues, a detailed investigation of the
radio/\tm spectral shape as a function of the radio spectral index
variations across the remnant is underway.  The association of
magnetic fields with the local R-T instabilities opens the door
for investigating the accompanying amplification of magnetic fields
in the remnant, and the roles of both relativistic particles and
fields in the further dynamical evolution of the remnant.

\acknowledgments

We appreciate the \tm images provided by C. Gerardy and R. Fesen, and
a number of important discussions on the science with Tom Jones. 
Support for work on Cas~A at the University of Minnesota was provided
by National Science Foundation grant AST-96-19438 and for comparisons
with HST data by NASA through grant number HST-AR-090537.01-A from the
Space Telescope Science Institute, which is operated by the Association of
Universities for Research in Astronomy, Inc., under NASA contract
NAS5-26555.  The comments of the referee were very helpful in improving
the paper.

\clearpage
\begin{deluxetable}{lrrrrrrc}
\label{tab:spectab}
\tablecolumns{8} 
\tablewidth{0pc} 
\tablecaption{Intensity and Polarization Measurements} 
\tablehead{ 
\colhead{ }    & Box 1 & Box 2    & Box 3  & Box 4  & Box 5 & Box 6 & 
Units (Error) \\}
\startdata
 &  &  &  &  &  &  &  \\
Azimuth &-10 & -20 & -30 & -35 & -42 & -32 & degrees  \\
$I_{2\mu m}$ observed & 12.0  & 18.6  & 7.6 & & 13.5  & 12.8 & $\mu$Jy/$\sq \arcsec$ (0.26) \\
$I_{2\mu m}$ corrected &  11.0  &  9.6  &  7.6  &   &  9.3  &  12.8 & 
$\mu$Jy /$\sq \arcsec$ (0.26)   \\
$I_{2\mu m}$ de-reddened &  18.3 & 15.9  & 12.6  &  & 15.4 & 21.2 & $\mu$Jy/$\sq \arcsec$ 
(0.43)\\
$Q'_{2\mu m}$ observed & -0.67 & -0.48 & -0.54 & -0.84 & -0.84 & -0.48 & 
$\mu$Jy/$\sq \arcsec$ (0.2) \\
$Q'_{2\mu m}$ de-reddened & -1.11 & -0.80  & -0.90 & -1.4 & -1.4 & -0.8 &  
$\mu$Jy/$\sq \arcsec$ (0.3) \\
$U'_{2\mu m}$ observed &-0.34 & .041 & 0.33 & 0.42  & 0.50  & 1.0  &  
$\mu$Jy/$\sq \arcsec$ (0.2) \\
$U'_{2\mu m}$ de-reddened &-0.56 & .068 & 0.55 & 0.70  & 0.83 & 1.66 &  
$\mu$Jy/$\sq \arcsec$ (0.3) \\
\%  $P_{2\mu m}$ & 6.2 & 3.8  & 7.2  &  & 9.9 & 8.3 & (1.5) \\
$PA_{{2\mu}} $ & 70 & 55 & 41 & 44 & 42 & 57 & degrees (13)\\
$I_{20cm}$ &  56,800 & 53,800 & 63,900 & 56,300 & 63,300 & 61,600 & 
$\mu$Jy/$\sq \arcsec$ (300) \\
$I_{6cm}$ &  21,500 & 20,400 & 24,200 & 21,500 & 23,500 & 23,200  & 
$\mu$Jy/$\sq \arcsec$ (100) \\
$Q_{6cm}$ &  -144  & 51  & 258  & 780  & 1535  & 163  & $\mu$Jy/$\sq \arcsec$ (45) \\
$U_{6cm}$ &  1400  & 965  & 1550  & 1270  & 700  & 680 & $\mu$Jy/$\sq \arcsec$ (45)  \\
\%  $P_{6cm}$ & 6.5 & 4.7  & 6.5 & 6.9 & 7.2 & 3.0 & (0.2) \\
$PA_{6cm}$ &  48 & 43.5 & 40.3 & 29.2 & 12.2 & 38.2 & degrees (2)  \\
$\approx$PA (filaments) & -40 & -40 & -33 & -70 & -78 & -33 & degrees (15) 
\\
 &  &  &  &  &  &  & \\
COMPARISONS &  & & & & & &\\
 &  &  &  &  &  &  & \\
\%  $P_{2.2\mu m}/\%  P_{6cm}$& 0.95 & 0.81 & 1.11 &  & 1.38 & 2.8 & (0.2)\\
$PA_{6cm}(corr) - PA_{2\mu}$ & -10 & -1.5 & +11 & -3 & -18 & -7 & 
degrees (13)\\
$\alpha^{6cm}_{20cm}$ & -0.76 & -0.75 & -0.76 & -0.75 & -0.77 & -0.76 & (.01)\\
$\alpha^{2.2\mu m}_{6cm}$ & -0.69 & -0.70 & -0.73 &  & -0.71 & -0.68 &  
($<.01$)\\
\enddata
\vskip 0.18in
Note: The absolute position angle (PA) on the sky at 
 \tm is $(0.5 \times tan^{-1}\frac{U'}{Q'} - 33)$ degrees.
 In the top section, the 6 cm PAs are as observed, {\it not} 
corrected for Faraday rotation. For comparison with the \tm PAs in
the bottom section, a correction for a rotation
 of $-106~ {\rm rad~m^{-2}}$ has been applied.
 \end{deluxetable}

\clearpage 

\epsscale{.60}
\plotone{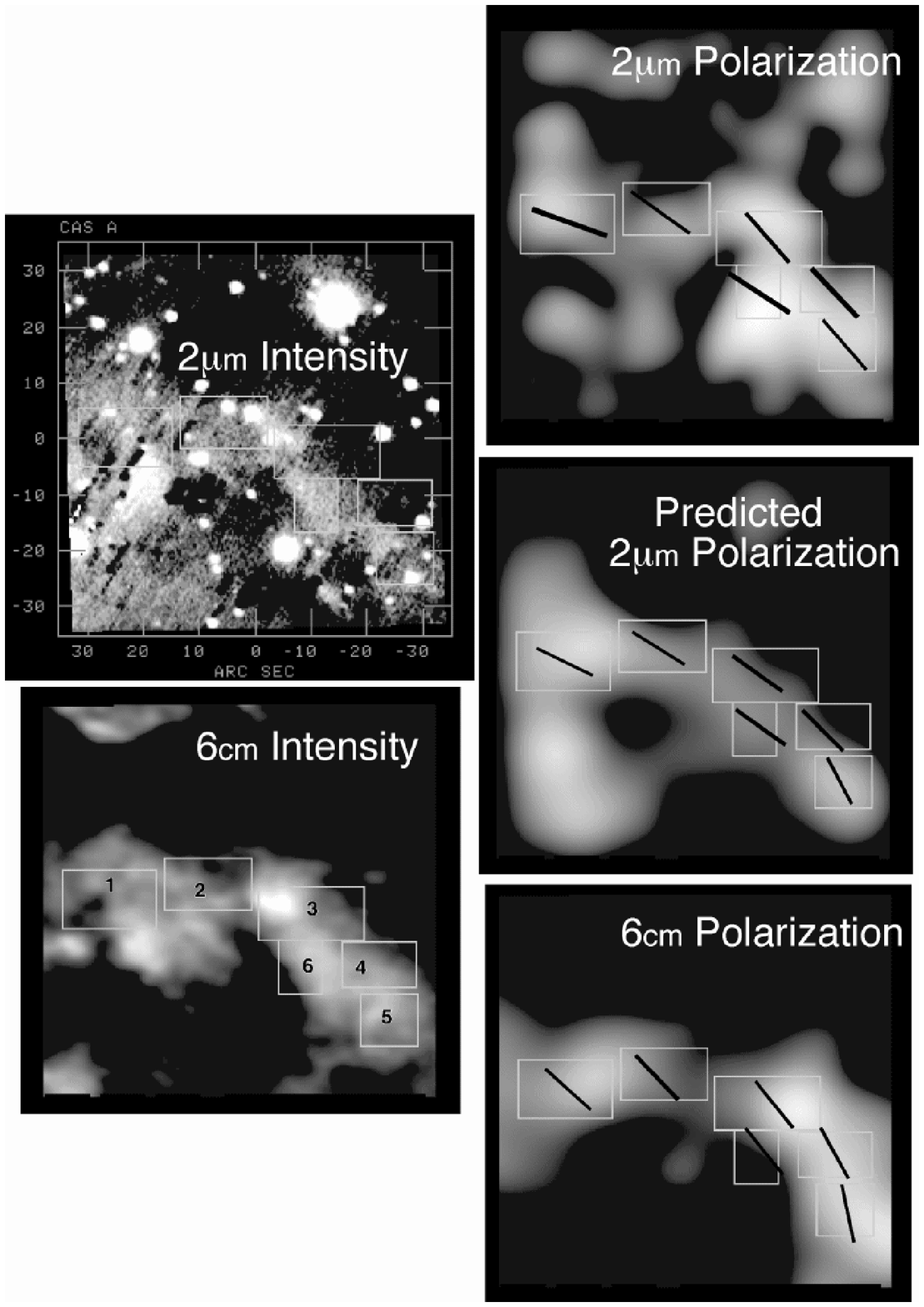}
\figcaption{\label{fig:total} Total intensity and polarization for 
Cassiopeia A at \tm and 6 cm. The left side images of total
intensity are at resolutions of 1$\arcsec$ (approximate seeing) and 
1.5$\arcsec$, respectively.
The center of all images is at RA 23h 23m 26s , 
Dec 58$^{\circ}$ 48$\arcmin$ 58.4$\arcsec$, (J2000),
and the  scale is shown on the \tm  map.  The images on the right 
are of polarized flux,  convolved to 10$\arcsec$,
along with the angle of the  {\it electric} field (observed PA) for
the boxes listed in Table 1.  The middle
frame on the right shows the predicted 2.2~ $\mu$m polarization 
predicted from the radio measurements, as described in the text.}

\clearpage

\plotone{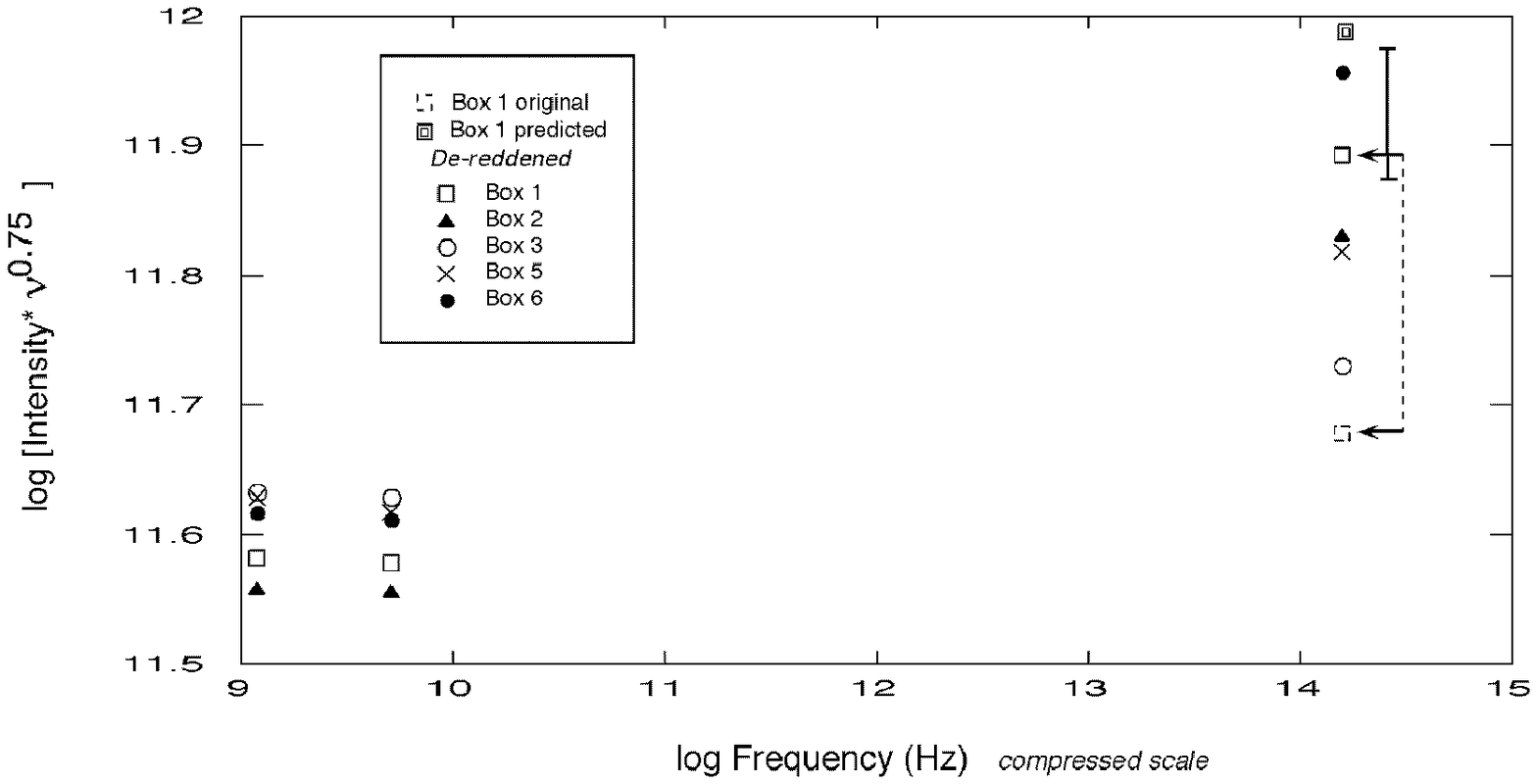}

\figcaption{\label{fig:fluxes} Flux densities for five of the boxes shown in 
Table 1, as a function of frequency.  The flux densities have been multiplied 
by a factor of $\nu^{0.75}$ to accentuate the curvature. The errors in
the radio are smaller than the symbol sizes. The errors at 2.2$\mu$m
are dominated by uncertainties in the extinction correction.
 The open and 
closed squares and the dashed line connecting them show the size of the
 nominal 
extinction correction applied to each box.  The error bar on the 
right shows the range of possible extinction corrections that could 
have been applied to the square (Box 1),
 as per the text discussion. The embedded
squares show a prediction for Box 1  based on non-linear acceleration at 
modified shocks, again as described in the text.}
\clearpage
\plotone{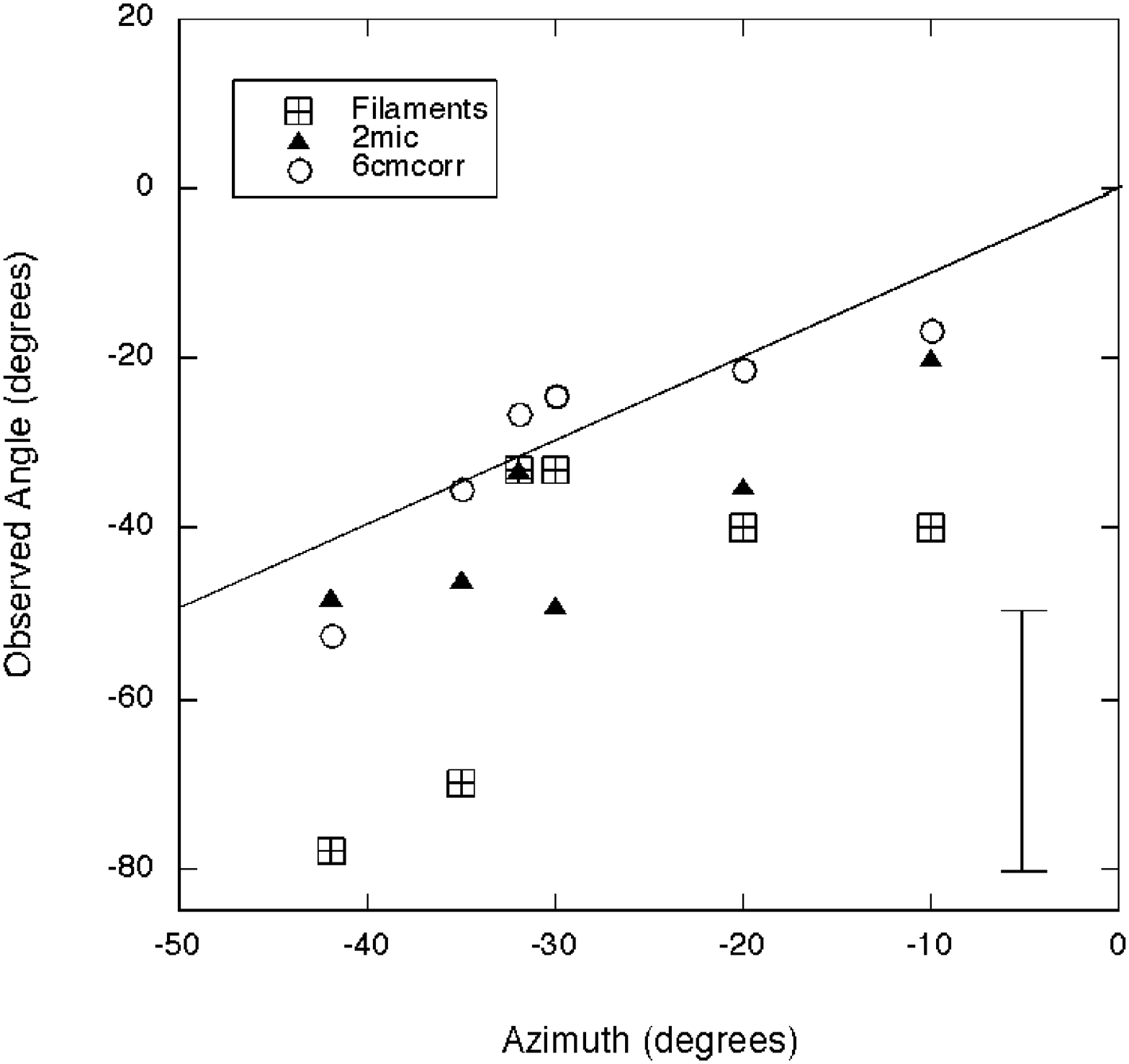}
\figcaption{\label{fig:angles} Inferred magnetic field angles (observed
PA + 90 $\deg$) and filament position 
angles as a function of azimuth.  A characteristic error is shown for the
2.2~ $\mu$m polarization and optical filament angles. 
The 6 cm angles have been corrected for the average RM in this region 
of -106 rad/m$^{2}$ and have errors of approximately the symbol size.
  The line indicates pure radial alignment.}

\end{document}